\documentclass[
aps,prl
 amsmath,amssymb,
 reprint,%
]{revtex4-1}
\usepackage{chngcntr} 
\usepackage{graphicx}
\usepackage{float}
\usepackage[font=small]{caption,subcaption}
\usepackage{makecell}
\usepackage{tabularx}
\usepackage{dcolumn}
\usepackage{bm}
\usepackage{tikz}
\usepackage[mathlines]{lineno}
\usepackage{mathptmx}
\usepackage{color}
\usepackage[colorlinks,linkcolor=blue]{hyperref}
\usepackage{physics}

\newcolumntype{Y}{>{\centering\arraybackslash}X}

\begin{document}
\title{Real-time equilibrium reconstruction by neural network based on HL-3 tokamak}

\author{G.H. Zheng$^{1, 2}$, S.F. Liu$^{2*}$, Z.Y. Yang$^{1}$, R. Ma$^{1}$, X.W. Gong$^{1}$, A. Wang$^{1}$, S. Wang$^{1}$, W.L. Zhong$^{1}$\\
\small{$^1$Southwestern Institute of Physics, Chengdu 610041,
People's Republic of China}\\
\small{$^2$School of Physics, Nankai University, Tianjin
300071, People's Republic of China}\\
\small{*Email : lsfnku@nankai.edu.cn}\\
}

\begin{abstract}
    A neural network model, EFITNN, has been developed capable of real-time magnetic equilibrium reconstruction based on HL-3 tokamak magnetic measurement signals. The model processes inputs from 68 channels of magnetic measurement data gathered from 1159 HL-3 experimental discharges, including plasma current, loop voltage, and the poloidal magnetic fields measured by equilibrium probes. The outputs of the model feature eight key plasma parameters, alongside high-resolution ($129\times129$) reconstructions of the toroidal current density $J_{\text P}$ and poloidal magnetic flux profiles $\Psi_{rz}$. Moreover, the network's architecture employs a multi-task learning structure, which enables the sharing of weights and mutual correction among different outputs, and lead to increase the model's accuracy by up to 32\%. The performance of EFITNN demonstrates remarkable consistency with the offline EFIT, achieving average $R^2 =$ 0.941, 0.997 and 0.959 for eight plasma parameters, $\Psi_{rz}$ and $J_{\text P}$, respectively. The model's robust generalization capabilities are particularly evident in its successful predictions of quasi-snowflake (QSF) divertor configurations and its adept handling of data from shot numbers or plasma current intervals not previously encountered during training. Compared to numerical methods, EFITNN significantly enhances computational efficiency with average computation time ranging from 0.08ms to 0.45ms, indicating its potential utility in real-time isoflux control and plasma profile management.

    \quad

    Keywords: HL-3 tokamak, real-time magnetic equilibrium reconstruction, neural network, EFITNN

\end{abstract}

\maketitle

\section{INTRODUCTION}
The reconstruction of magnetic equilibria by various magnetic measurement data collected during discharges is fundamental in analyzing plasma states and enhancing the performance of tokamak discharges. Through the analysis of data from diverse probes and coils, it's possible to ascertain the plasma's geometry, safety factor profile, distributions of magnetic flux and current density, etc. These determinations are significant for advancing plasma physics research, exploring operational modes, and developing diagnostic techniques. On this basis, the realization of real-time plasma configuration control becomes viable with the reduction of magnetic flux profile computation times to the millisecond level. Additionally, the accurate and rapid calculation of plasma parameters is a prerequisite for the implementation of various burgeoning algorithms, such as real-time disruption warning systems. Therefore, it is essential to develop an algorithm capable of accurately and swiftly reconstructing plasma parameters during the discharge process of a tokamak. The EFIT (Equilibrium reconstruction and fitting) code \cite{lao1985reconstruction} is extensively adopted for its precision and practicality. Currently, EFIT or similar equilibrium reconstruction systems are employed in tokamaks across the globe, including DIII-D, JET, NSTX, EAST, KSTAR, START, C-MOD, TORE SUPRA, HL-2A, QUEST, MAST, among others \cite{lao2005mhd, o1992equilibrium, sabbagh2001equilibrium, jinping2009equilibrium, li2013kinetic, park2011kstar, jiang2021kinetic, appel2001equilibrium, appel2006unified, in2000resistive, zwingmann2003equilibrium, li2011efit, hongda2006study, xue2019equilibrium, berkery2021kinetic}.

The fundamental principle of the offline EFIT code is solving the ideal magnetohydrodynamic equilibrium equation under poloidal axisymmetry, known as the Grad-Shafranov (GS) equation \cite{grad1958hydromagnetic, shafranov1966plasma}. Typically, the computation for a single time slice spans several seconds, which falls short of the requirements of real-time control applications. To bridge this gap and facilitate equilibrium reconstruction for real-time tokamak control, extensive efforts have been made to improve and optimize the EFIT code. Innovations include the development of a real-time EFIT (rt-EFIT) numerical algorithm based on EFIT \cite{rui2018acceleration, ferron1998real}, which achieves millisecond-scale computation speeds. Additionally, efforts on the EAST tokamak have led to the acceleration of the EFIT process by utilizing GPUs (P-EFIT) \cite{huang2020gpu}, significantly improving the convergence rate of the algorithm. Nonetheless, these advancements often necessitate a compromise on lower resolution of reconstructions or a presumption of gradual changes in magnetic surfaces. Therefore, these works always lead to various performance trade-offs in practice.

To overcome the computation limitations inherent in the offline EFIT program while maintaining necessary accuracy, machine learning (ML) presents an compelling solution. Machine learning is developed to enable automated learning of characteristics and patterns from extensive existing datasets, thereby allowing machines to predict unseen data based on these learned patterns. Currently, the fusion research community has placed significantly emphasis on the application of machine learning, including instability identification \cite{fu2020machine, akccay2021machine}, disruption prediction \cite{ferreira2019deep, guo2021disruption, yang2019disruption, rea2018disruption, montes2019machine, kates2019predicting, piccione2020physics}, recognition and prediction of high confinement mode (H mode) \cite{jacobus2022machine, gaudio2014alternative}, and the identification and categorization of various instabilities \cite{fu2020machine, akccay2021machine, li2022simulation, hui2021machine}. Moreover, machine learning has proven effective as a rapid numerical solver \cite{mathews2021uncovering, citrin2015real, felici2018real, van2020fast}, capable of reconstructing plasma profiles and the last closed flux surface (LCFS) \cite{matos2017deep, wan2023machine}, and facilitating real-time control of plasma configurations during discharges \cite{degrave2022magnetic}.

For machine learning in magnetic equilibrium reconstruction, efforts have been made on the KSTAR to employ deep learning as a solver for GS equation \cite{joung2019deep}. In this work, a network is designed with inputs consist of real-time magnetic measurements and a targeted coordinate, and the output is the magnetic flux value at that location. However, its network structure is relatively simple and does not fully utilize some conditions hidden in the offline EFIT principle. There is also a neural network model trained on EFIT data from the DIII-D device \cite{lao2022application} for real-time reconstruction of the magnetic flux profile. Besides, another model with predictions of parameters such as safty factor $q_{95}$ and the normalized plasma pressure $\beta_N$ is also examined. Similar attempts have also been made on the Globus-M2 and NSTX-U tokamaks \cite{mitrishkin2021new, wai2022neural}. Furthermore, the EFIT neural network model based on the EAST device fine-tunes various hyperparameters by the Optuna framework \cite{lu2023fast}, achieving high prediction accuracy. These studies demonstrate that machine learning methods can perform rapid magnetic equilibrium reconstruction with high precision. However, the aforementioned works mainly focus on the preditions of poloidal magnetic flux distribution and a limited number of plasma parameters. They could be improved by paying more attention to the design of the model structure, and taking advantage of principles concealed in the EFIT code. 

In this paper, a neural network named EFITNN has been developed for real-time magnetic equilibrium reconstruction of plasma. Compared to the models mentioned above, EFITNN is capable of predicting eight plasma parameters to support plasma shape control and feedback control of parameters such as $\beta_{\text{P}}$ and safety factor. Additionally, EFITNN outputs high-resolution ($129\times129$) profiles of poloidal magnetic flux and toroidal current density. The network benefits from a multi-task output structure, where different outputs share part network layers and weights. Different branches can foster mutual correction and complementation among each other, potentially enhancing the model's overall accuracy. Furthermore, EFITNN leverages the foundational principles of the EFIT algorithm, which acknowledges the local correlations in the spatial distribution of magnetic flux and other profiles. Compared to fully connected layers typically employed in general networks, EFITNN is better suited to identify local features within the data by incorporating deconvolution and convolution layers into the network. The data for model training and validation is derived from discharges of HL-3, along with results from the offline EFIT. HL-3 is a new tokamak device constructed by the Southwestern Institute of Physics in China, designed to support a maximum plasma current capacity of 2.5-3MA and a total heating power of up to 27MW \cite{duan2022progress}. Recently, HL-3 achieved its inaugural H-mode discharge with a current of up to 1MA. EFITNN has shown promising predictive capabilities with the experimental data from HL-3 discharges, demonstrating its potential as a tool for enhancing real-time control of the plasma.

The rest of this paper is organized as follows: Section 2 introduces the data collected for model and outlines the data preprocessing steps. Follow this, Section 3 elucidates the architecture and hyperparameters of EFITNN. This section also includes a comparative analysis to highlight the benefits of employing the multi-task learning approach. Section 4 assesses the model's predictive and outinference performance, focusing on the reconstruction of 8 plasma parameters, $\Psi_{rz}$ and $J_{\text P}$ profiles. Finally, we summarize our work and provide a perspective on future research in section 5.

\quad

\section{Data collection and preprocessing}

\begin{figure*}[t]
    \centering
    \begin{subfigure}{.365\textwidth}
        \includegraphics[width=\textwidth, height=1.428\textwidth]{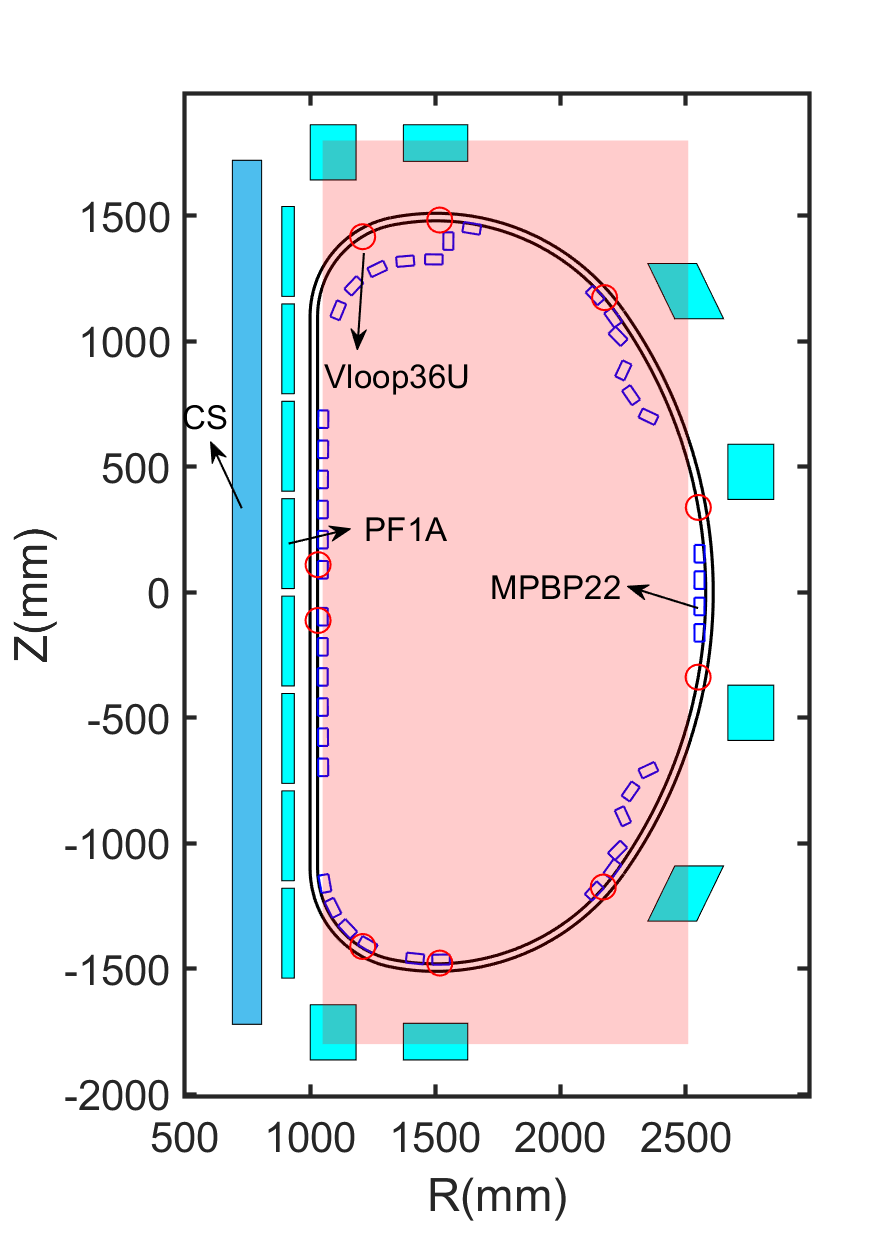}
        \caption*{(a)}
    \end{subfigure}
    \quad
    \begin{subfigure}{.589\textwidth}
        \includegraphics[width=\textwidth, height=.833\textwidth]{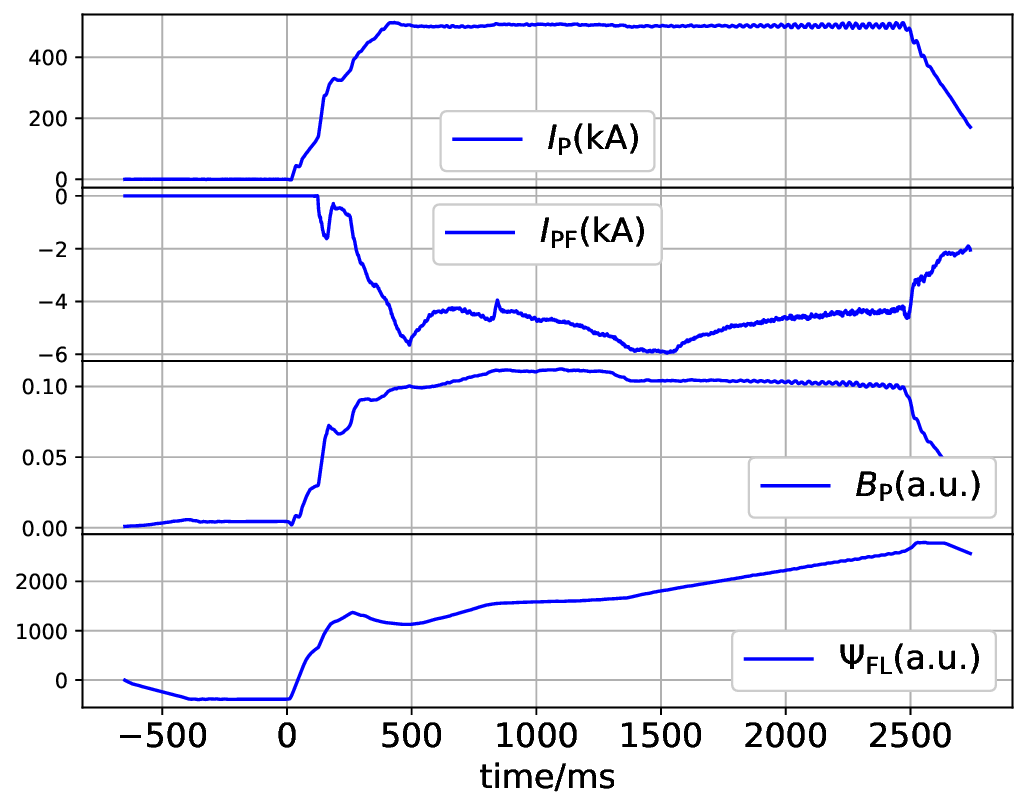}
        \caption*{(b)}
    \end{subfigure}
    \caption{(a) Locations of probes for the poloidal magnetic field and loop voltage measurements on HL-3. The red circles represent the positions of 10 loop voltage measurement devices, and the blue square points indicate the locations of 41 poloidal magnetic field probes (only 39 of them are used in EFITNN). The blue and purple rectangulars correspond to the PF and CS coils, respectively. The pink rectangular area represents the region covered by the magnetic flux and current density distribution maps. (b) Curves from top to bottom are the plasma current, PF coil current (selected from PF1A), poloidal magnetic field (selected from MPBP22), and the intergration of the loop voltage (proportional to the poloidal magnetic flux, selected from VLoop36U). These data are collected from shot \#3331.}
    \label{fig:profile}
\end{figure*}

The dataset for training and evaluating the network comprises data from HL-3 stable discharges, spanning from shot \#1248 to shot \#4186, totaling 1159 shots with 1 515 618 time slices. These time slices are proportionately divided based on the shot number into a training set (shots \#1248-\#3308, 70\%), a validation set (shots \#3309-\#4010, 20\%), and a test set (shots \#4011-\#4186, 10\%). The training set is used to facilitate the learning of data features and to minimize error rates, while the validation set serves to evaluate the model's learning efficacy after each training epoch. The model reaches the lowest loss on the validation set will be chosen as the best. However, this approach implies that the validation set contributes indirectly to the model's refinement. The true predictive prowess of the network is gauged by the test set, to ascertain the model's performance on entirely unknown data. Strategically segmenting the dataset into training, validation, and test sets in chronological order of shot numbers mirrors the experimental application scenario, because the discharge configurations always change gradually as the shot number increasing. By this means, the network's predictive performance will be well-aligned under a similarly real-time environment.

The input for the network consists of 68 channels of experimental measurements, which covers plasma current, loop voltage, poloidal magnetic fields from equilibrium probes, and the currents from the PF (Poloidal Field), TF (Toroidal Field) and CS (Central Solenoid) coils. These inputs are carefully selected to be consistent with those of the offline EFIT, but with the additional capability of being collected in real-time. Figure \ref{fig:profile}(a), (b) respectively showcase the probe locations for these measurements and the temporal variation of the input data for a representative shot (\#3331). 

\begin{figure}
    \centering
    \includegraphics[width=.9\linewidth]{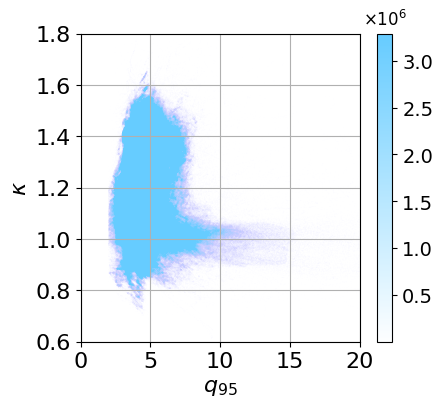}
    \caption{The relationship between plasma elongation $\kappa$ (vertical axis) and the safety factor at 95\% of the plasma minor radius $q_{95}$ (horizontal axis). The data are derived from shots \#1248-\#4186, and the color indicates the density of points.}
    \label{fig:eq_distribution}
\end{figure}

The network's outputs are comprehensive, including eight key plasma parameters and detailed $129\times129$ grid values of $\Psi_{rz}$ and $J_{\text P}$ distributions. A summary of these inputs and outputs, with their respectively names, meanings, and channel numbers is provided in table \ref{tab:params}. Additionally, figure \ref{fig:eq_distribution} offers a visual representation of the dataset, displaying the relationship between the elongation $\kappa$ and safety factor $q_{95}$ across all available time slices. The performance of the network proves that the selection of inputs is well-equipped to handle the intricate reconstructions, with a high degree of precision.

\begin{table}
    \centering
    \caption{Summary of the data used in the network.}
    \begin{tabular}{lll}
        \Xhline{1pt}
        \Gape[8pt]{Name} & Definition & No. \\
        \hline
        \Gape[6pt]{Inputs} & \quad & 68 \\
        \hline
        \Gape[6pt]{$\Psi_{\text{FL}}$} & The intergration of the loop voltage & 10 \\
        \Gape[6pt]{$B_{\text P}$} & Poloidal magnetic field & 39 \\
        \Gape[6pt]{$I_{\text{PF}}$} & Poloidal field coil current & 16 \\
        \Gape[6pt]{$I_\text{TF}$} & poloidal field coil current & 1 \\
        \Gape[6pt]{$I_\text{CS}$} & Central solenoid coil current & 1 \\
        \Gape[6pt]{$I_\text{P}$} & Plasma current & 1 \\
        \hline
        \Gape[6pt]{Outputs} & \quad & 33290 \\
        \hline
        \Gape[6pt]{$a$} & Plasma minor radius & 1 \\
        \Gape[6pt]{$\kappa$} & Plasma elongation & 1 \\
        \Gape[6pt]{$l_i$} & Plasma inductance & 1 \\
        \Gape[6pt]{$q_{95}$} & Safe factor in 95\% minor radius & 1 \\
        \Gape[6pt]{$r_C$} & \makecell[l]{Coordinate of plasma geometric \\center in $r$} & 1 \\
        \Gape[6pt]{$z_C$} & \makecell[l]{Coordinate of plasma geometric \\center in $z$} & 1 \\
        \Gape[6pt]{$\delta_\text B$} & Bottom triangularity of the plasma & 1 \\
        \Gape[6pt]{$\delta_\text T$} & Top triangularity of the plasma & 1 \\
        \Gape[6pt]{$\Psi_{rz}$} & Poloidal magnetic flux distribution & 16641 \\
        \Gape[6pt]{$J_{\text p}$} & Toroidal current density distribution & 16641 \\
        \Xhline{1pt}
    \end{tabular}
    \label{tab:params}
\end{table}

Prior to training, all data underwent Z-score normalization by channel. This process is essential for mitigating disparities in data magnitude across various channels, thereby facilitating a more efficient learning process. The normalization formula is represented as follows:
\begin{equation}
    x' = \dfrac{x - \bar x}{\sigma_x},
\end{equation}

\noindent where $x$ is the original value, $x'$ is the normalized value, $\bar x$ is the mean value of the data for a given channel, and $\sigma_x$ is the standard deviation of the data for a specific channel. Notably, the parameters for normalization ($\bar x$ and $\sigma_x$) are calculated exclusively from the training set, which ensures that the validation and test sets remain unaffected by the training data.

Data truncation is an additional preprocessing step applied to each shot within the dataset to address the significant fluctuations in plasma current, which are typically observed at the beginning and end of the discharge. These fluctuations can induce complex eddy currents on the vacuum chamber walls, which may distort the measurements from magnetic probes and potentially result in inaccurate EFIT calculations. To mitigate this issue, time slices from the initial and final phases of the discharge, where the plasma current is not stable, are excluded from the dataset. The criteria for data inclusion are as follows: only time slices for the plasmas current higher than 80kA are considered. Moreover, all subsequent time slices are disregarded in the event of a disruption during the discharge. By setting a threshold for the plasma current and narrowing the time window, the dataset can be ensured representative of the stable phase of the discharge.

\section{Model structures and parameters tuning}

\subsection{Network layers}
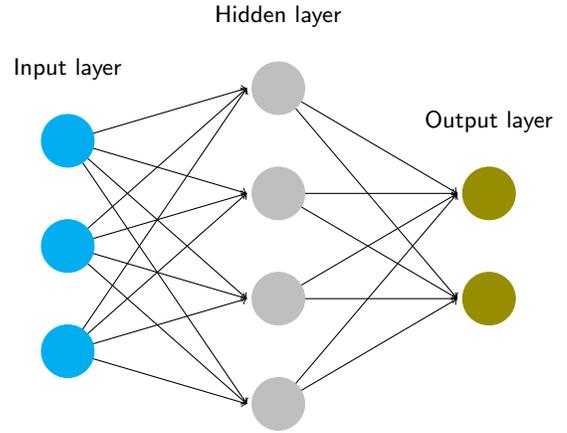
\begin{figure}
    \centering
    \begin{tikzpicture}[scale=0.7]

        \node [above, black] at (0, 6) {\textsf{Input layer}};
        \node [above, black] at (4, 7) {\textsf{Hidden layer}};
        \node [above, black] at (8, 5) {\textsf{Output layer}};

        \draw[<->] (3.4, 0) -- (0, 1) -- (3.4, 6);
        \draw[<->] (3.4, 2) -- (0, 1) -- (3.4, 4);
        \draw[<->] (3.4, 0) -- (0, 3) -- (3.4, 6);
        \draw[<->] (3.4, 2) -- (0, 3) -- (3.4, 4);
        \draw[<->] (3.4, 0) -- (0, 5) -- (3.4, 6);
        \draw[<->] (3.4, 2) -- (0, 5) -- (3.4, 4);

        \draw[<->] (7.4, 4) -- (4, 6) -- (7.4, 2);
        \draw[<->] (7.4, 4) -- (4, 4) -- (7.4, 2);
        \draw[<->] (7.4, 4) -- (4, 2) -- (7.4, 2);
        \draw[<->] (7.4, 4) -- (4, 0) -- (7.4, 2);

        \draw[cyan, fill=cyan] (0, 1) circle [radius=0.5];
        \draw[cyan, fill=cyan] (0, 3) circle [radius=0.5];
        \draw[cyan, fill=cyan] (0, 5) circle [radius=0.5];

        \draw[lightgray, fill=lightgray] (4, 0) circle [radius=0.5];
        \draw[lightgray, fill=lightgray] (4, 2) circle [radius=0.5];
        \draw[lightgray, fill=lightgray] (4, 4) circle [radius=0.5];
        \draw[lightgray, fill=lightgray] (4, 6) circle [radius=0.5];

        \draw[olive, fill=olive] (8, 4) circle [radius=0.5];
        \draw[olive, fill=olive] (8, 2) circle [radius=0.5];

    \end{tikzpicture}
    \caption{A fully-connected network with 3 layers.}
    \label{fig:fully-connected}
\end{figure}

EFITNN is constructed based on the principles of artificial neural networks (ANNs), which are computational models inspired by the human brain's neural system. ANNs are adept at learning from data and structured to include an input layer, multiple hidden layers, and an output layer. The most prevalent form of ANNs is the fully connected neural network, as shown in figure \ref{fig:fully-connected}.

In a fully connected neural network, each node, or neuron, in a given layer $l$ is connected to every node in the adjacent layer through a set of weights. For a particular node $a_j^l$ in layer $l$, its value is computed as the weighted sum of all nodes in the preceding layer $l-1$, plus a bias term $b_j^l$. This relationship is mathematically expressed as follows:
\begin{equation}
    a_j^l = f \left ( \sum_i w_{ij}a_i^{l-1} + b_j^l \right ).
\end{equation}

\noindent In this equation, $w_{ij}$ represents the weight connecting node $a_i^{l-1}$ in layer $l-1$ to node $a_j^l$ in layer $l$, and $b_j^l$ is the bias associated with node $a_j^l$. Both $w_{ij}$ and $b_j^l$ are parameters that the network learns during the training process. The function $f$ is known as the activation function, which introduces non-linearity into the model, allowing it to capture complex patterns in the data. Common choices for activation functions include $\tanh$ and $\text{ReLU}$. For EFITNN, the Gaussian Error Linear Unit (GeLU) function has been chosen as the activation function due to its properties and performance in deep learning tasks, which can be expressed as

\begin{equation}
    \text{GeLU}(x) = \frac x2 \cdot \text{Erf} \left( \frac {x} {\sqrt{2}} \right),
\end{equation}

\noindent where $\text{Erf}(x)$ is the Gaussian Error Function.

It is well-established that a three-layer neural network with a sufficient number of neurons and appropriate weights can approximate any continuous function to an arbitrary degree of accuracy \cite{cybenko1989approximation}. However, in practice, the complexity of a neural network must be carefully managed to avoid overfitting, especially when the available training samples are limited. Overfitting will significantly impair the model's ability to generalize to new, unseen data.

Therefore, to make efficient use of the available weights and nodes without overly complicating the network structure, innovative locally connected network architectures have been introduced. Convolutional neural networks (CNNs) \cite{lecun1989backpropagation} are one such development, renowned for their proficiency in recognizing local and intricate features within data. Correspondingly, deconvolutional networks \cite{zeiler2014visualizing, khan2018guide} have been designed to reconstruct detailed information from compressed inputs. 

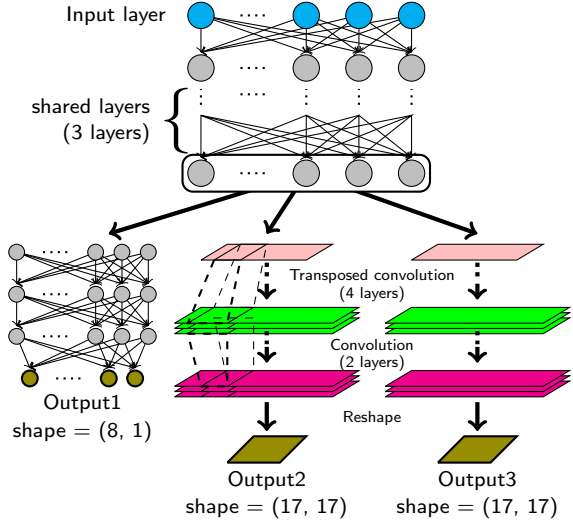
\begin{figure}
    \centering

    \begin{tikzpicture}[scale=0.35]
        \draw [<->] (4, -0.4) -- (-4, 1) -- (-4, -0.4);
        \draw [<->] (4, -0.4) -- (0, 1) -- (-4, -0.4);
        \draw [<->] (4, -0.4) -- (2, 1) -- (-4, -0.4);
        \draw [<->] (4, -0.4) -- (4, 1) -- (-4, -0.4);

        \draw [<->] (0, -0.4) -- (-4, 1) -- (2, -0.4);
        \draw [<->] (0, -0.4) -- (0, 1) -- (2, -0.4);
        \draw [<->] (0, -0.4) -- (2, 1) -- (2, -0.4);
        \draw [<->] (0, -0.4) -- (4, 1) -- (2, -0.4);

        \node [left] at (-5, 1) {\fontsize{8}{24}\selectfont \textsf {Input layer}};
        \draw [fill=cyan] (-4, 1) circle [radius=0.5];
        \draw [dotted, thick] (-2.5, 1) -- (-1.5, 1);
        \draw [fill=cyan] (0, 1) circle [radius=0.5];
        \draw [fill=cyan] (2, 1) circle [radius=0.5];
        \draw [fill=cyan] (4, 1) circle [radius=0.5];

        \node [left] at (-4, -3) {\fontsize{35}{24}\selectfont\textsf{\{}};
        \node [left] at (-5.6, -2.5) {\fontsize{8}{24}\selectfont \textsf {shared layers}};
        \node [left] at (-5.6, -3.5) {\fontsize{8}{24}\selectfont \textsf {(3 layers)}};

        \draw [fill=lightgray] (-4, -1) circle [radius=0.5];
        \draw [dotted, thick] (-2.5, -1) -- (-1.5, -1);
        \draw [fill=lightgray] (0, -1) circle [radius=0.5];
        \draw [fill=lightgray] (2, -1) circle [radius=0.5];
        \draw [fill=lightgray] (4, -1) circle [radius=0.5];

        \draw [dotted, thick] (-4, -2.4) -- (-4, -1.8);
        \draw [dotted, thick] (-2.5, -2) -- (-1.5, -2);
        \draw [dotted, thick] (0, -2.4) -- (0, -1.8);
        \draw [dotted, thick] (2, -2.4) -- (2, -1.8);
        \draw [dotted, thick] (4, -2.4) -- (4, -1.8);

        \draw [ultra thick, ->] (0, -5) -- (-7.5, -7.3);
        \draw [ultra thick, ->] (0, -5) -- (-1.5, -7.3);
        \draw [ultra thick, ->] (0, -5) -- (6.5, -7.3);
        \draw [thick, rounded corners, fill=white] (-4.7, -5.7) rectangle (4.7, -4.3);

        \draw [<->] (4, -4.4) -- (-4, -2.8) -- (-4, -4.4);
        \draw [<->] (4, -4.4) -- (0, -2.8) -- (-4, -4.4);
        \draw [<->] (4, -4.4) -- (2, -2.8) -- (-4, -4.4);
        \draw [<->] (4, -4.4) -- (4, -2.8) -- (-4, -4.4);

        \draw [<->] (0, -4.4) -- (-4, -2.8) -- (2, -4.4);
        \draw [<->] (0, -4.4) -- (0, -2.8) -- (2, -4.4);
        \draw [<->] (0, -4.4) -- (2, -2.8) -- (2, -4.4);
        \draw [<->] (0, -4.4) -- (4, -2.8) -- (2, -4.4);
        
        \draw [fill=lightgray] (-4, -5) circle [radius=0.5];
        \draw [dotted, thick] (-2.5, -5) -- (-1.5, -5);
        \draw [fill=lightgray] (0, -5) circle [radius=0.5];
        \draw [fill=lightgray] (2, -5) circle [radius=0.5];
        \draw [fill=lightgray] (4, -5) circle [radius=0.5];

        \draw [<->] (-11, -9.2) -- (-11, -8) -- (-6, -9.2);
        \draw [<->] (-11, -9.2) -- (-8, -8) -- (-6, -9.2);
        \draw [<->] (-11, -9.2) -- (-7, -8) -- (-6, -9.2);
        \draw [<->] (-11, -9.2) -- (-6, -8) -- (-6, -9.2);

        \draw [<->] (-8, -9.2) -- (-11, -8) -- (-7, -9.2);
        \draw [<->] (-8, -9.2) -- (-8, -8) -- (-7, -9.2);
        \draw [<->] (-8, -9.2) -- (-7, -8) -- (-7, -9.2);
        \draw [<->] (-8, -9.2) -- (-6, -8) -- (-7, -9.2);

        \draw [fill=lightgray] (-11, -8) circle [radius=0.3];
        \draw [dotted, thick] (-10, -8) -- (-9, -8);
        \draw [fill=lightgray] (-8, -8) circle [radius=0.3];
        \draw [fill=lightgray] (-7, -8) circle [radius=0.3];
        \draw [fill=lightgray] (-6, -8) circle [radius=0.3];

        \draw [<->] (-11, -10.8) -- (-11, -9.6) -- (-6, -10.8);
        \draw [<->] (-11, -10.8) -- (-8, -9.6) -- (-6, -10.8);
        \draw [<->] (-11, -10.8) -- (-7, -9.6) -- (-6, -10.8);
        \draw [<->] (-11, -10.8) -- (-6, -9.6) -- (-6, -10.8);

        \draw [<->] (-8, -10.8) -- (-11, -9.6) -- (-7, -10.8);
        \draw [<->] (-8, -10.8) -- (-8, -9.6) -- (-7, -10.8);
        \draw [<->] (-8, -10.8) -- (-7, -9.6) -- (-7, -10.8);
        \draw [<->] (-8, -10.8) -- (-6, -9.6) -- (-7, -10.8);

        \draw [fill=lightgray] (-11, -9.6) circle [radius=0.3];
        \draw [dotted, thick] (-10, -9.6) -- (-9, -9.6);
        \draw [fill=lightgray] (-8, -9.6) circle [radius=0.3];
        \draw [fill=lightgray] (-7, -9.6) circle [radius=0.3];
        \draw [fill=lightgray] (-6, -9.6) circle [radius=0.3];

        \draw [<->] (-10.5, -12.4) -- (-11, -11.2) -- (-6.5, -12.4);
        \draw [<->] (-10.5, -12.4) -- (-8, -11.2) -- (-6.5, -12.4);
        \draw [<->] (-10.5, -12.4) -- (-7, -11.2) -- (-6.5, -12.4);
        \draw [<->] (-10.5, -12.4) -- (-6, -11.2) -- (-6.5, -12.4);

        \draw [->] (-11, -11.2) -- (-7.5, -12.4);
        \draw [->] (-8, -11.2) -- (-7.5, -12.4);
        \draw [->] (-7, -11.2) -- (-7.5, -12.4);
        \draw [->] (-6, -11.2) -- (-7.5, -12.4);

        \draw [fill=lightgray] (-11, -11.2) circle [radius=0.3];
        \draw [dotted, thick] (-10, -11.2) -- (-9, -11.2);
        \draw [fill=lightgray] (-8, -11.2) circle [radius=0.3];
        \draw [fill=lightgray] (-7, -11.2) circle [radius=0.3];
        \draw [fill=lightgray] (-6, -11.2) circle [radius=0.3];

        \draw [thick, fill=olive] (-10.5, -12.8) circle [radius=0.3];
        \draw [dotted, thick] (-9.5, -12.8) -- (-8.5, -12.8);
        \draw [thick, fill=olive] (-7.5, -12.8) circle [radius=0.3];
        \draw [thick, fill=olive] (-6.5, -12.8) circle [radius=0.3];

        \node [below] at (-8.5, -13.1) {\fontsize{8}{24}\selectfont \textsf {Output1}};
        \node [below] at (-8.5, -14.1) {\fontsize{8}{24}\selectfont \textsf {shape = (8, 1)}};

        \draw [fill=pink] (-4, -8.3) -- (0, -8.3) -- (1, -7.7) -- (-3, -7.7) -- (-4, -8.3);
        \draw (-3.5, -8.3) -- (-2.5, -8.3) -- (-1.5, -7.7) -- (-2.5, -7.7) -- (-3.5, -8.3);

        \draw [fill=green] (-5, -11.1) -- (1, -11.1) -- (2, -10.5) -- (-4, -10.5) -- (-5, -11.1);
        \draw (-4.5, -11.1) -- (-3.5, -10.5);
        \draw (-3, -11.1) -- (-2, -10.5);
        \draw [fill=green] (-5, -10.9) -- (1, -10.9) -- (2, -10.3) -- (-4, -10.3) -- (-5, -10.9);
        \draw (-4.5, -10.9) -- (-3.5, -10.3);
        \draw (-3, -10.9) -- (-2, -10.3);
        \draw [fill=green] (-5, -10.7) -- (1, -10.7) -- (2, -10.1) -- (-4, -10.1) -- (-5, -10.7);
        \draw [fill=green] (-4.5, -10.7) -- (-3, -10.7) -- (-2, -10.1) -- (-3.5, -10.1) -- (-4.5, -10.7);
        
        \draw [dashed, thick] (-3.5, -8.3) -- (-4.5, -10.7) -- (-3, -10.7) -- (-2.5, -8.3);
        \draw [dashed] (-2.5, -7.7) -- (-3.5, -10.1) -- (-2, -10.1) -- (-1.5, -7.7);
        
        \draw [fill=magenta] (-5, -13.5) -- (1, -13.5) -- (2, -12.9) -- (-4, -12.9) -- (-5, -13.5);
        \draw (-4, -13.5) -- (-3, -12.9);
        \draw (-3, -13.5) -- (-2, -12.9);
        \draw [fill=magenta] (-5, -13.3) -- (1, -13.3) -- (2, -12.7) -- (-4, -12.7) -- (-5, -13.3);
        \draw (-4, -13.3) -- (-3, -12.7);
        \draw (-3, -13.3) -- (-2, -12.7);
        \draw [fill=magenta] (-5, -13.1) -- (1, -13.1) -- (2, -12.5) -- (-4, -12.5) -- (-5, -13.1);
        \draw (-4, -13.1) -- (-3, -12.5);
        \draw (-3, -13.1) -- (-2, -12.5);

        \draw [dashed, thick] (-4, -13.1) -- (-3, -13.1) -- (-3, -11.1) -- (-4.5, -11.1) -- (-4, -13.1);
        \draw [dashed] (-3, -12.5) -- (-2, -12.5) -- (-2, -10.5) -- (-3.5, -10.5) -- (-3, -12.9);

        \draw [fill=olive, thick] (-3, -16) -- (-1, -16) -- (0, -15) -- (-2, -15) -- (-3, -16);

        \draw [dotted, ultra thick] (-1.5, -8.4) -- (-1.5, -9);
        \draw [->, ultra thick] (-1.5, -9.2) -- (-1.5, -9.8);

        \draw [dotted, ultra thick] (-1.5, -11) -- (-1.5, -11.5);
        \draw [->, ultra thick] (-1.5, -11.6) -- (-1.5, -12.2);

        \draw [->, ultra thick] (-1.5, -13.7) -- (-1.5, -14.8);

        \node [below] at (-1.5, -16) {\fontsize{8}{24}\selectfont \textsf {Output2}};
        \node [below] at (-1.5, -17) {\fontsize{8}{24}\selectfont \textsf {shape = (17, 17)}};


        \draw [fill=pink] (4, -8.3) -- (8, -8.3) -- (9, -7.7) -- (5, -7.7) -- (4, -8.3);
        
        \draw [fill=green] (3, -11.1) -- (9, -11.1) -- (10, -10.5) -- (4, -10.5) -- (3, -11.1);
        \draw [fill=green] (3, -10.9) -- (9, -10.9) -- (10, -10.3) -- (4, -10.3) -- (3, -10.9);
        \draw [fill=green] (3, -10.7) -- (9, -10.7) -- (10, -10.1) -- (4, -10.1) -- (3, -10.7);
        
        \draw [fill=magenta] (3, -13.5) -- (9, -13.5) -- (10, -12.9) -- (4, -12.9) -- (3, -13.5);
        \draw [fill=magenta] (3, -13.3) -- (9, -13.3) -- (10, -12.7) -- (4, -12.7) -- (3, -13.3);
        \draw [fill=magenta] (3, -13.1) -- (9, -13.1) -- (10, -12.5) -- (4, -12.5) -- (3, -13.1);



        \draw [fill=olive, thick] (5, -16) -- (7, -16) -- (8, -15) -- (6, -15) -- (5, -16);

        \draw [dotted, ultra thick] (6.5, -8.4) -- (6.5, -9);
        \draw [->, ultra thick] (6.5, -9.2) -- (6.5, -9.8);

        \draw [dotted, ultra thick] (6.5, -11) -- (6.5, -11.5);
        \draw [->, ultra thick] (6.5, -11.6) -- (6.5, -12.2);

        \draw [->, ultra thick] (6.5, -13.7) -- (6.5, -14.8);

        \node [below] at (6.5, -16) {\fontsize{8}{24}\selectfont \textsf {Output3}};
        \node [below] at (6.5, -17) {\fontsize{8}{24}\selectfont \textsf {shape = (17, 17)}};

        \node at (2.5, -8.9) {\fontsize{6}{24}\selectfont \textsf {Transposed convolution}};
        \node at (2.5, -9.6) {\fontsize{6}{24}\selectfont \textsf {(4 layers)}};

        \node at (2.5, -11.5) {\fontsize{6}{24}\selectfont \textsf {Convolution}};
        \node at (2.5, -12.2) {\fontsize{6}{24}\selectfont \textsf {(2 layers)}};

        \node at (2.5, -14.3) {\fontsize{6}{24}\selectfont \textsf {Reshape}};

    \end{tikzpicture}

    \caption{Structure of the model mainly for predicting plasma parameters (three outputs). The input layer receives 68 channels of real-time magnetic measurement data, while output layers include 8 plasma parameters, $17\times17$ grids of $\Psi_{rz}$ and $J_{\text P}$. The low-resolution outputs for magnetic flux and current density are solely to assist in the training of plasma parameters and not considered as the final reconstruction results. When the network primarily objective is to reconstruct $129\times 129$ magnetic flux or current density, strides and kernel sizes of the convolution and deconvolution layers need to be adjusted to fit the new output dimensions. Comprehensive details about the hyperparameters of convolution and deconvolution layers for each specific objective are provided in appendix A.}
    \label{fig:structure}
\end{figure}

EFITNN incorporates a combination of these network layers. The shared layers and those predicting plasma parameters are composed excluded of fully connected layers, while the layers tasked with predicting $\Psi_{rz}$ and $J_{\text P}$ also integrate deconvolution and convolution layers to enhance the network's ability to capture spatial details. Figure \ref{fig:structure} provides an schematic representation of the network's architecture and the interplay between its constituent layers.

\subsection{Loss functions and back propagations}
To quantify the network's ability to fit the data, a measurable criterion known as the loss function is essential. For EFITNN, the mean square error (MSE) is employed as the loss function. It is defined as
\begin{equation}
    \mathcal{L} \equiv \frac1{N} \sum_i^N \left |\mathbf{y_i} - \mathbf{\hat y_i} \right |^2,
\end{equation}

\noindent where $N$ represents the total sample number, $\mathbf{y_i}$ denotes the $i$th true value, and $\mathbf{\hat y_i}$ is the $i$th predicted value by the model. The objective of training the model is to minimize the loss by continuously updating the weights and biases between nodes, referring to as back propagation of error. The optimization algorithm used to achieve this minimization is known as the optimizer. One of the foundational optimizers is the Stochastic Gradient Descent (SGD) \cite{robbins1951stochastic}. Within a fully connected neural network, the SGD updates to weights and biases as follows:
\begin{equation}
    w_{ji} \leftarrow w_{ji} - \eta \dfrac{\partial}{\partial w_{ji}} \mathcal{L}(w_{ji}),
\end{equation}

\begin{equation}
    b_j \leftarrow b_j - \eta \dfrac{\partial }{\partial b_j} \mathcal{L}(b_j).
\end{equation}

\noindent Here, $\eta$ is known as the learning rate, a critical hyperparameter in the neural network training process. Building upon the SGD, various advanced optimizers have been developed, such as SGD with momentum, Adam and AdamW. \cite{polyak1964some, kingma2014adam,loshchilov2017decoupled}. These optimizers share the core principle of SGD but incorporate additional features, such as regularization techniques or more refined adjustments to the weight updates. For EFITNN, the AdamW optimizer is selected, since AdamW can not only perform the gradient descent, but also integrate a term known as weight decay into the weight update equation as a regularization process. The updated formula for weights is
\begin{equation}
    w_{ji} \leftarrow w_{ji} - \eta \left (\dfrac{\partial}{\partial w_{ji}} \mathcal{L}(w_{ji}) + \alpha w_{ji}\right ).
\end{equation}

Here, $\alpha$ is introduced as an additional hyperparameter, referred to as weight decay. This simple mechanism is particularly effective in preventing overfitting and thus preserving the network's generalization capabilities.

\subsection{Model structures and training}

\begin{figure*}[t]
    \centering
    \begin{subfigure}{.48\textwidth}
        \includegraphics[width=\textwidth]{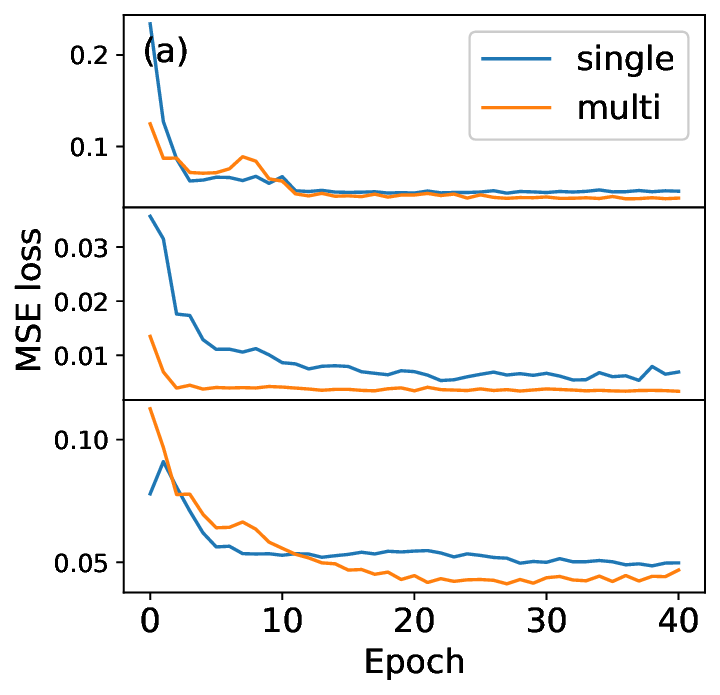}
    \end{subfigure}
    \quad
    \begin{subfigure}{.46\textwidth}
        \includegraphics[width=\textwidth]{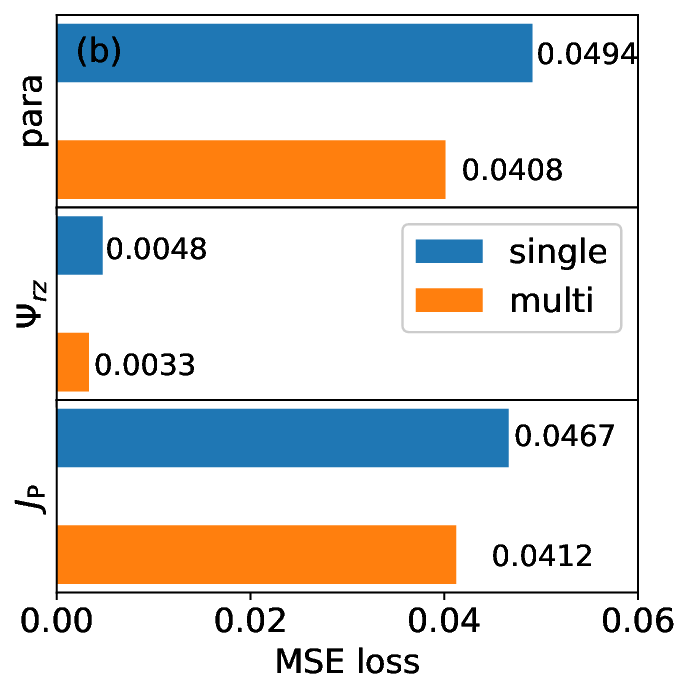}

    \end{subfigure}
    \caption{Comparison of validation set losses between single output and multi-task learning structure during the training process. (a) From top to bottom, the losses of models mainly predicting para, $\Psi_{rz}$, and $J_{p}$ over training epochs. (b) The minimum validation loss values of the models.}
    \label{fig:multitask}
\end{figure*}

\begin{figure*}[t]
    \centering
    \includegraphics[width=\textwidth]{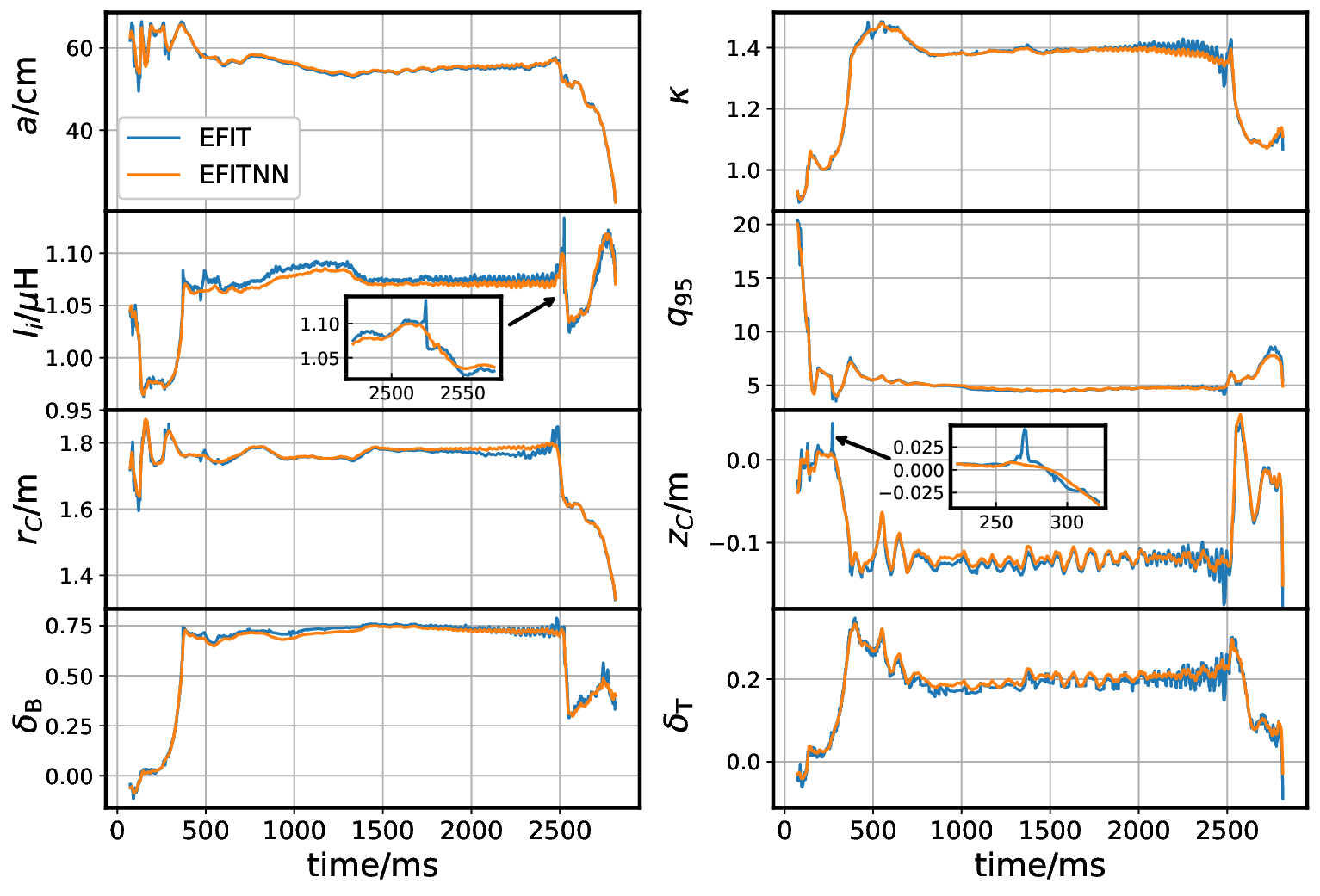}
    \caption{Curves of the EFITNN and offline EFIT values for 8 plasma parameters over time. The data is derived from shot \#3331. The figure also includes detailed views for $l_i$ and $z_C$, which indicates that the model can correct obviously erroneous data compared with offline EFIT, demonstrating good generalization capability of the model.}
    \label{fig:para_pred}
\end{figure*}

EFITNN is designed as a multi-task learning network, which means that the model features a single input layer that feeds into multiple output layers. The structure of EFITNN primarily focused on predicting plasma parameters is displayed in figure \ref{fig:structure}.

All outputs are inherently interconnected under the same set of physical laws, suggesting that the results across the outputs would exhibit a degree of correlations. By employing a multi-output architecture, the model implicitly incorporates the constraints of these physical mechanisms into its structure. Therefore, the different branches of the network can share some layers to correct each other. For instance, if one branch's outputs are trapped in a local minimum during the gradient descent, the updates to the shared weight layers from the other branches may facilitate an escape from the local minimum. Notably, in the offline EFIT, the distribution of $\Psi_{rz}$ is a prerequisite for calculating eight plasma parameters. Consequently, the branch dedicated to predicting the $\Psi_{rz}$ distribution likely plays a significant role in enhancing the accuracy of the plasma parameter predictions. 

To empirically demonstrate the multi-task learning models outperform their single-output counterparts, we have conducted a comparative study of the training effects between them. The comparison focuses on the loss values of 8 plasma parameters, $\Psi_{rz}$ and $J_{\text P}$ distributions, as depicted in figure \ref{fig:multitask}.

The results of this comparison highlight the superior performance of multi-task learning over single-task learning. The adoption of multi-task learning models leads to a significant reduction in the validation set loss values. Models predicting plasma parameters, $\Psi_{rz}$ and $J_{\text P}$ show obviously decreases in the loss value with 17\%, 32\%, and 12\%, respectively.

In a multi-task learning network structure, each output is associated with its own loss function. The final loss used to update the weights from shared layers is a composite of the individual losses from each output, with each loss being assigned a specific weight coefficient. Therefore, the loss weight of the primary training objective is typically assigned higher compared to the auxiliary outputs. After evaluating the model's performance and the resources consumed during training, EFITNN has been configured with different multi-task learning structures tailored for three objectives. Their outputs and corresponding loss weights are presented in table \ref{tab:structure}.

\begin{table}
    \centering
    \caption{Network structures for predicting plasma parameters (para), $\Psi_{rz}$ and $J_{p}$. To reduce computational expense and memory usage, the outputs for $\Psi_{rz}$ and $J_{\text P}$ in the network mainly predicting plasma parameters have a resolution of $17\times17$, while in other networks, the resolution is $129\times129$.}
    \begin{tabularx}{.45\textwidth}{lYY}
        \Xhline{1pt}
        \Gape[8pt]{Target} & Outputs & Weights \\
        \hline
        \Gape[6pt]{para} & (para, $\Psi_{rz}$, $J_{p}$) & (1.0, 0.3, 0.1)\\
        \Gape[6pt]{$\Psi_{rz}$} & (para, $\Psi_{rz}$) & (0.2, 1.0) \\
        \Gape[6pt]{$J_{p}$} & (para, $J_{p}$) & (0.2, 1.0) \\
        \Xhline{1pt}
    \end{tabularx}
    \label{tab:structure}

\end{table}

The development of EFITNN has been conducted using the PyTorch framework, with some of common hyperparameters in table \ref{tab:config}.

\begin{table}
    \centering
    \caption{Some hyperparameters of the model. These hyperparameters are common across the networks for predicting plasma parameters, $\Psi_{rz}$ and $J_{\text P}$ distributions, while other hyperparameters are presented in appendix A. The learning rate and weight decay decrease gradually during the training process, and only their minimum values are listed in the table.}
    \begin{tabularx}{.45\textwidth}{lY}
        \Xhline{1pt}
        \Gape[8pt]{Definition} & Value \\
        \hline
        \Gape[6pt]{Learning rate} & $1.5\times 10^{-5}$ \\
        \Gape[6pt]{Weight decaying} & $1.5\times 10^{-4}$ \\
        \Gape[6pt]{Batch size} & 128 \\
        \Gape[6pt]{Activation function} & GeLU \\
        \Gape[6pt]{Optimizer} & AdamW \\
        \Gape[6pt]{Loss function} & MSE \\
        \Gape[6pt]{No. of Shared layers} & 3 \\
        \Gape[6pt]{No. of nodes in each shared layer} & 512 \\
        \Gape[6pt]{No. of para layers} & 3 \\
        \Gape[6pt]{No. of nodes in each para layer} & 512 \\
        \Xhline{1pt}
    \end{tabularx}
    \label{tab:config}

\end{table}

\section{Results and analysis}
\subsection{Prediction cability of models}

In this work, we assess the predictive performance of the network across all time slices within the validation and test sets. The evaluation metrics include the correlation coefficient $r$, coefficient of determination $R^2$, and the mean absolute value of the normalized residuals. The formulas for these calculations are performed as follows:

\begin{equation}
    r = \dfrac{\displaystyle \sum_i^N (y_i - \langle y \rangle)(\hat y_i - \langle \hat y \rangle)}{\displaystyle\sqrt{\sum_i^N(y_i - \langle y \rangle)^2 \sum_i^N(\hat y_i - \langle \hat y \rangle)^2}},
\end{equation}

\begin{equation}
    R^2 = 1 - \dfrac{\displaystyle\sum_i^N (y_i - \hat y)^2}{\displaystyle\sum_i^N (y_i - \langle y \rangle)^2},
\end{equation}

\begin{equation}
    \left |\Delta \right |= \dfrac1N \sum_i^N \left |\dfrac{y_i - \hat{y_i}}{\max(|y_i|)} \right | .
\end{equation}

\noindent The results examined on validation and test sets are compiled in table \ref{tab:para_pred}, with their correlation figures shown in appendix B.

\begin{table}
    \centering
    \caption{Correlation coefficient $r$, coefficient of determination $R^2$, and the mean absolute value of the normalized residuals $|\Delta|$ between the predicted and the EFIT values of 8 plasma parameters, $\Psi_{rz}$, and $J_{\text P}$ for the validation and test sets (shots \#3309-\#4186). The above data represents the average of calculations performed separately for each shot. Detailed comparisons between results of EFITNN and  EFIT are displayed in appendix B.}
    \begin{tabularx}{.45\textwidth}{cYYY}
        \Xhline{1pt}
        \Gape[8pt]{Name} & $r$ & $R^2$ & $|\Delta|$ \\
        \hline
        \Gape[6pt]{$a$} & 0.976 & 0.939 & 0.0101\\
        \Gape[6pt]{$\kappa$} & 0.992 & 0.979 & 0.0087\\
        \Gape[6pt]{$l_i$} & 0.967 & 0.906 & 0.0069\\
        \Gape[6pt]{$q_{95}$} & 0.972 & 0.933 & 0.0108\\
        \Gape[6pt]{$r_C$} & 0.974 & 0.941 & 0.0047\\
        \Gape[6pt]{$z_C$} & 0.991 & 0.977 & 0.0287\\
        \Gape[6pt]{$\delta_B$} & 0.984 & 0.929 & 0.0409\\
        \Gape[6pt]{$\delta_T$} & 0.974 & 0.923 & 0.0561\\ 
        \Gape[6pt]{$\Psi_{rz}$} & 0.999 & 0.997 & 0.0081 \\
        \Gape[6pt]{$J_{\text P}$} & 0.981 & 0.959 & 0.0082 \\
        \Xhline{1pt}
    \end{tabularx}
    \label{tab:para_pred}
\end{table}

Figure \ref{fig:para_pred} presents the temporal evolution of the plasma parameters for shot \#3331, which is the first ELMy-H mode discharge in HL-3. This shot has not encountered during the model's training phase, so it will be an ideal candidate to test EFITNN's generalization capability.

\begin{figure}
    \centering
    \includegraphics[width=.48\textwidth]{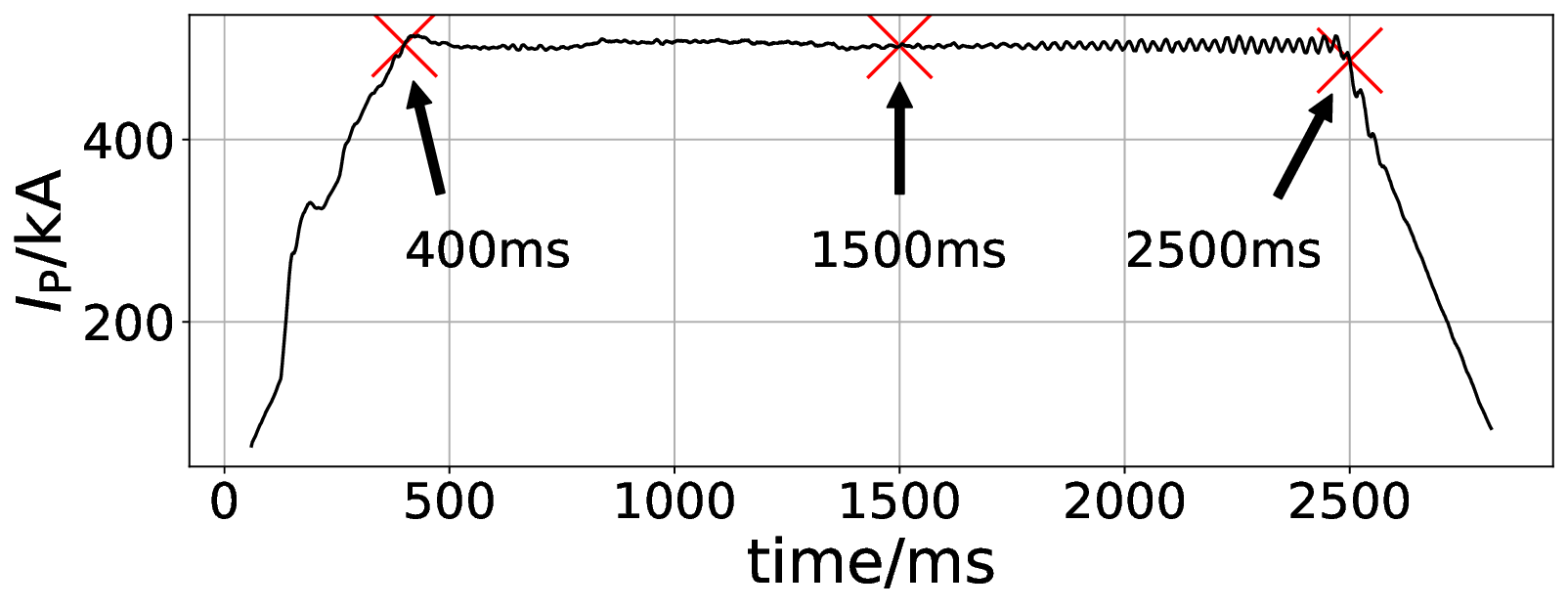}
    \caption{Time evolution of the plasma current for shot \#3331. The moments $\Psi_{rz}$ and $J_{\text P}$ plotted in figure \ref{fig:psirz} are marked in the curve.}
    \label{fig:Ip}
\end{figure}

\begin{figure*}[t]
    \centering
    \includegraphics[width=\textwidth]{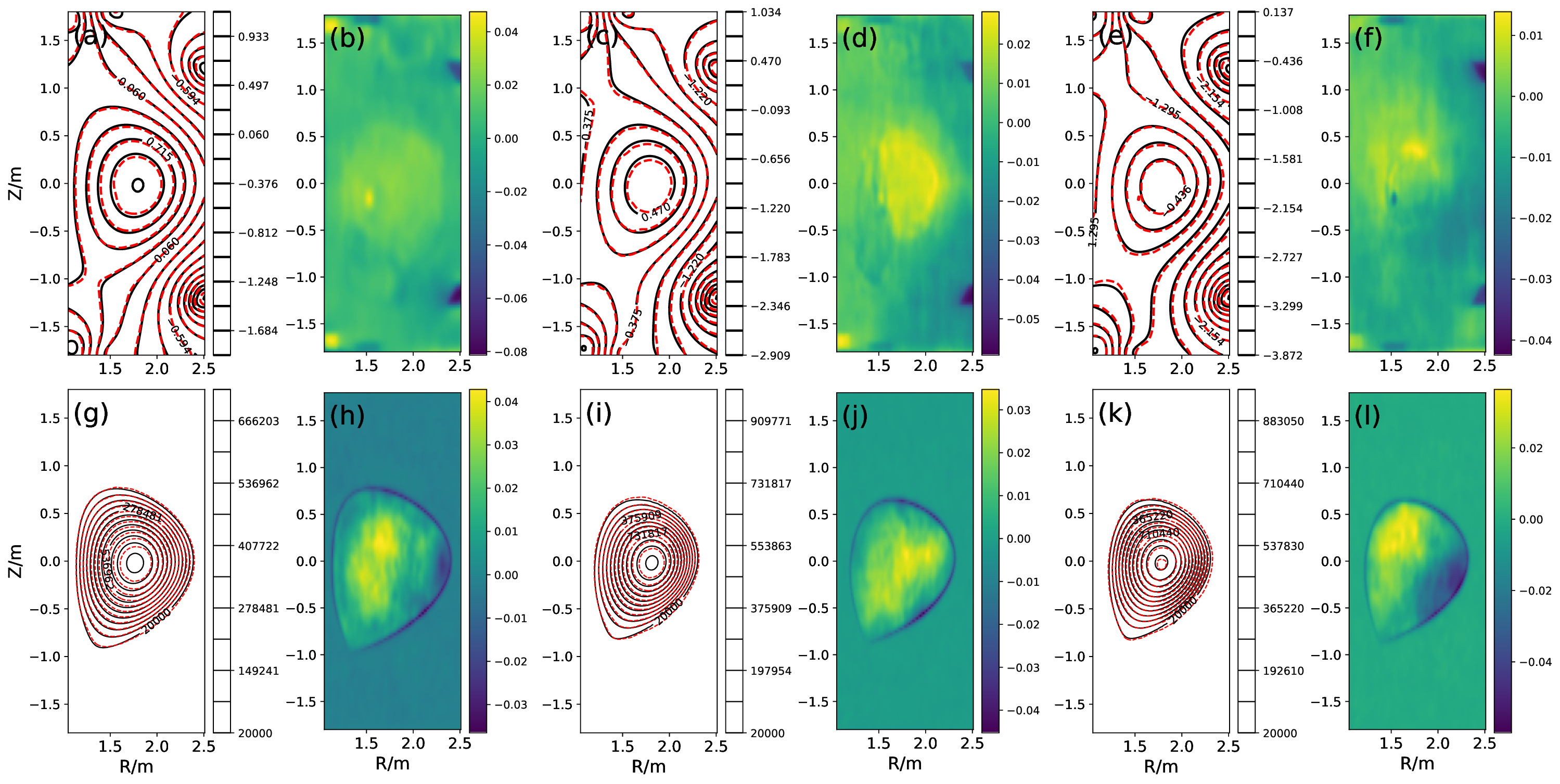}
    \caption{The model's predictive capability for $\Psi_{rz}$ and $J_{\text P}$ at three time points for shot \#3331 (as shown in figure \ref{fig:Ip}). Subfigures (a)(c)(e) compare the predicted (red dashed lines) and EFIT (black solid lines) values of $\Psi_{rz}$. Subfigures (b)(d)(f) depict the normalized residuals of $\Psi_{rz}$ at these time points. Subfigures (g)-(l) compare the $J_{\text P}$ distribution, following the same layout as subfigures (a)-(f).}
    \label{fig:psirz}
\end{figure*}

\begin{figure*}
    \centering
    \includegraphics[width=\textwidth]{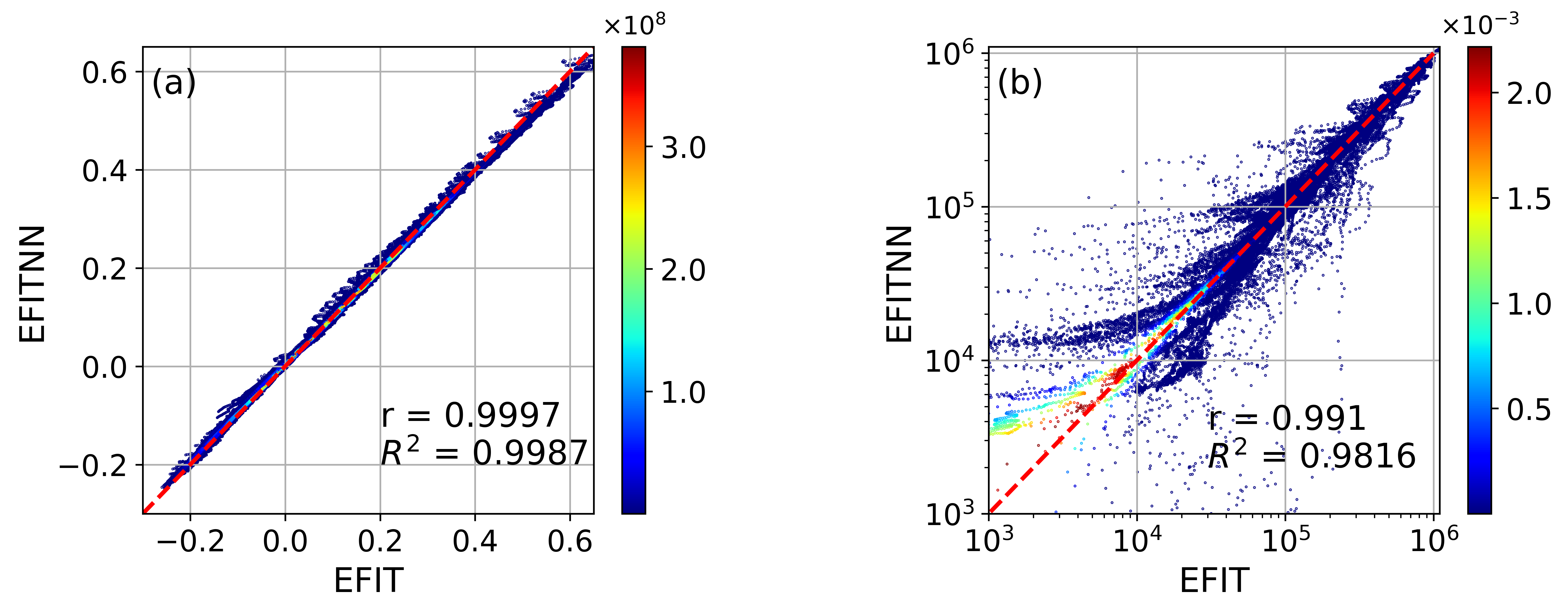}

    \caption{Comparison of values between offline EFIT and EFITNN for (a) $\Psi_{rz}$ and (b) $J_{\text P}$ in shot \#3331. Due to the large number of data points, a uniform sampling of the $129\times129$ distribution values has been performed for each moment when plotting the scatter plot, resulting in a $17\times17$ distribution for all moments of magnetic flux and current density distributions, totaling $2741\times17\times17=792149$ sample points. The red dashed line represents the $y=x$ line, and the color of the dots indicates the density of data points in the sample space.}
    \label{fig:r2}
\end{figure*}

An encouraging consistency can be discerned from the data illustrated in figure \ref{fig:para_pred}. It is worth mentioning that the offline EFIT code may sometimes produce discontinuities at certain moments, possibly due to insufficient computation time or saturation of the magnetic probes' current. The discontinuities from EFIT will result in abrupt peaks in time evolution curves. In contrast, EFITNN's predictions provide a smoother curve, which suggest that the model is capable of correcting these anomalies and offers a robust alternative to traditional methods.

A detailed evaluation is also conducted for the model's performance in predicting the distributions of $\Psi_{rz}$ and $J_{\text P}$. In this test, three specific time points during shot \#3331 were chosen, including the inflection points of current rise and fall, as well as a stable point. The plasma current curve and the selected time points are depicted in figure \ref{fig:Ip}. Figure \ref{fig:psirz} shows the comparison results between EFITNN predictions and offline EFIT for $\Psi_{rz}$ and $J_{\text P}$ at these moments, including heatmaps that visualize the normalized residuals. Additionally, figure \ref{fig:r2} provides the correlation plots between them.

\begin{figure}
    \centering
    \includegraphics[width=.45\textwidth]{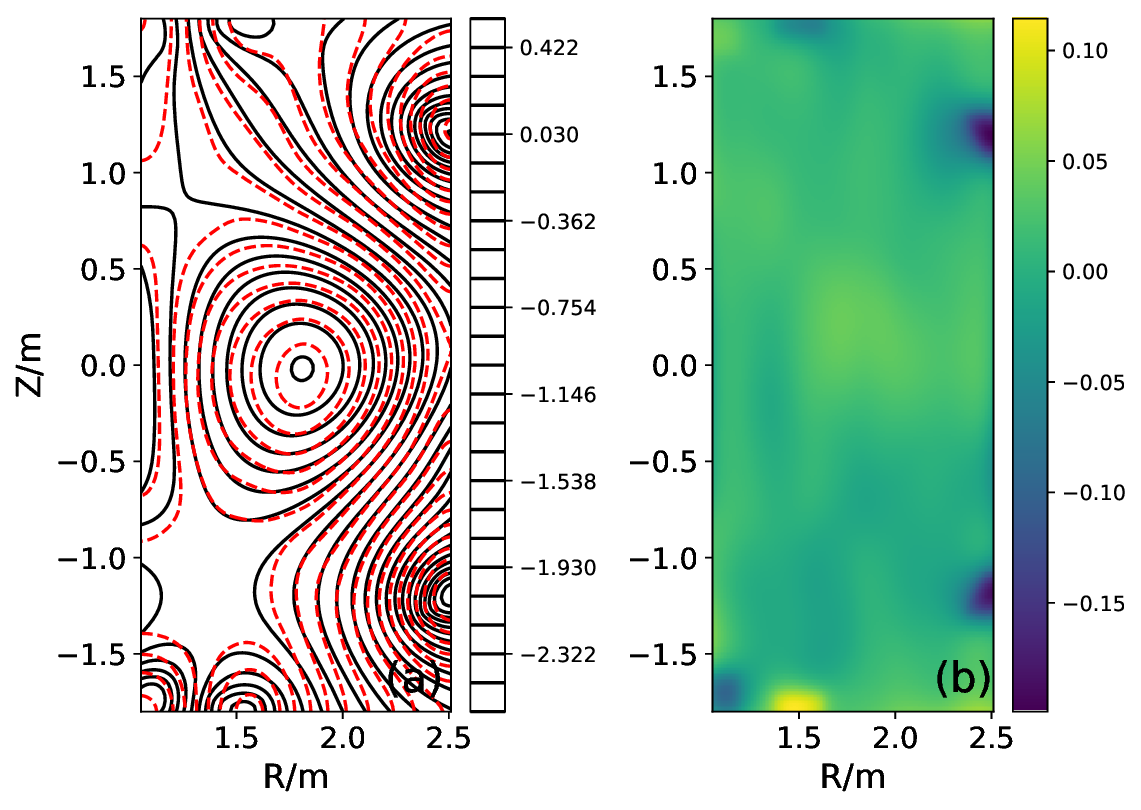}
    \caption{(a) EFITNN predicted values (red dashed line) and EFIT reference values (black solid line) of $\Psi_{rz}$ at 1100ms for shot \#4104; (b) Residual map for the predictions of $\Psi_{rz}$.}
    \label{fig:snow}
\end{figure}

It is noteworthy that within the latter portion of the test set, a small subset of discharges features QSF divertor configurations. These configurations present flux shapes that are markedly different from those seen in the training set. Despite with lower accuracy than normal profile distributions, EFITNN is able to successfully predict the $\Psi_{rz}$ profile for this shape. Figure \ref{fig:snow} showcases the predicted images and residual maps for the $\Psi_{rz}$ distribution of one such snowflake shape (Shot \#4104, 1100ms). The accomplishment underscores EFITNN's robust generalization capabilities, and also suggests that the 68 channels of magnetic measurement data chosen in this study are reasonable. The input data obtained from the channels appear to sufficiently capture the essential information required to determine the poloidal magnetic flux distribution of the plasma.

In general, EFITNN exhibits a high degree of agreement with the results obtained from the offline EFIT results. However, as indicated in figure \ref{fig:r2}, the model's predictive performance is somewhat less accurate in regions characterized by lower current density. The regions often coincide with phases of the discharge that involve rapid changes in current, such as during the ramp-up and ramp-down periods. As discussed in Section 2, the eddy currents induced in the vessel walls can affect the accuracy of the magnetic probes. Consequently, the results predicted by the model from these regions may be less reliable.

\subsection{Model extraploration performance on unseen parameter intervals}

The ability of a model to extrapolate and maintain high accuracy on data which has not been trained on is a critical aspect of its overall performance in machine learning. A robust model should not only excel with the training set but also demonstrate consistent accuracy on the novel data. In this part, we examine the extrapolation performance of the model by analyzing the coefficient of determination ($R^2$) for three different outputs across shots and plasma current intervals not included in the training set.

\begin{figure}[h]
    \centering
    \includegraphics[width=.48\textwidth]{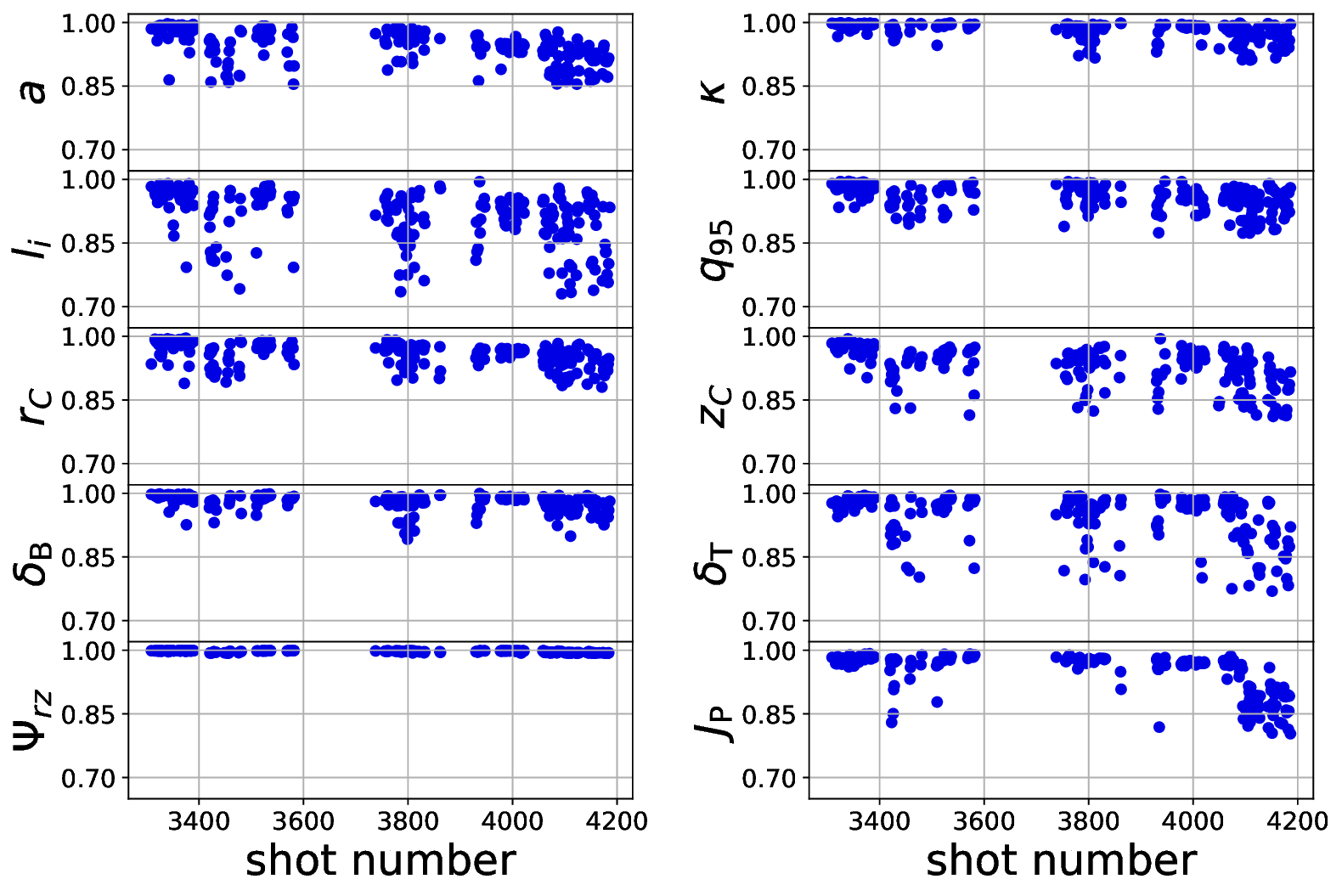}
    \caption{Illustration of the model's extrapolation capability for various outputs by shot number. The horizontal axis represents the increasing shot numbers in the validation and test sets, and the vertical axis represents the coefficient of determination $R^2$.}
    \label{fig:outinference}
\end{figure}

Figure \ref{fig:outinference} displays a scatter plot that correlates the shot number with the $R^2$ values. It is notable that while the training set data is randomized, the validation and test sets are sequentially organized by shot number and time. Consequently, the shot numbers on the x-axis of figure \ref{fig:outinference} progressively diverge from the range of the training set. The arrangement allows for an evaluation of the model's extrapolation capabilities by observing the trend of $R^2$ values as they relate to the shot number. The model's performance in predicting all outputs remains relatively stable, as depicted in figure \ref{fig:outinference}. However, there are still some variations in prediction accuracy for the latter part of the test set (particularly after shot \#4078), which may be attributed to significant changes for discharge parameters, such as the introduction of QSF configuration.

\begin{figure}[h]
    \centering
    \includegraphics[width=.48\textwidth]{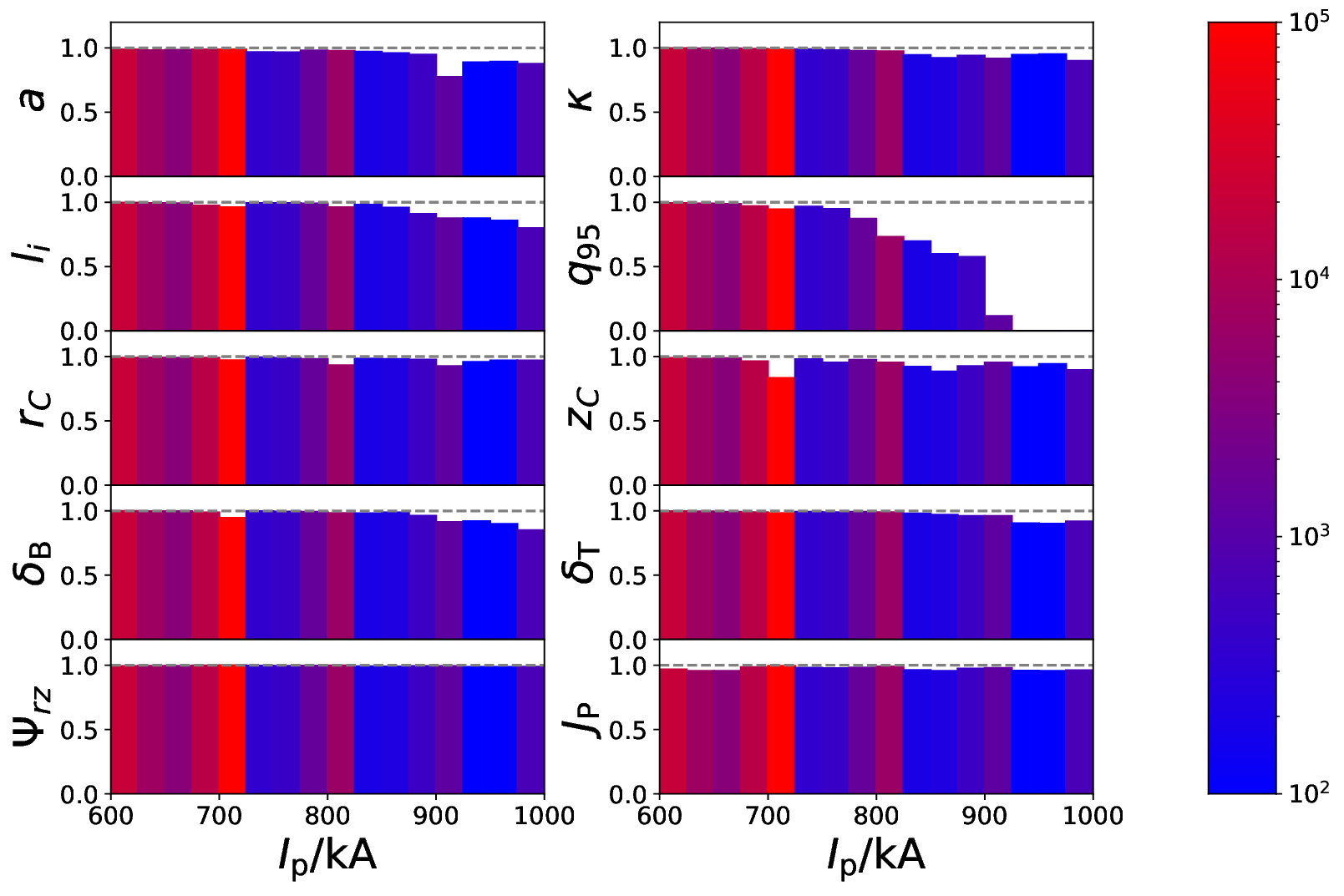}
    \caption{Variations of the model's extrapolation capability by plasma current. The horizontal axis represents the increasing intervals of plasma current; the vertical axis represents the $R^2$ within these intervals, and the color indicates the number of samples in each current interval.}
    \label{fig:outinference2}
\end{figure}

The assessment of a model's generalization capability often involves examining its performance across varying conditions, such as different magnitudes of plasma current. To facilitate the analysis, the training and validation sets are reorganized based on plasma current values. Specifically, data with plasma currents below 600kA are designated as the new training set (about 80\%), while time slices with plasma currents ranging from 600kA to 1000kA are allocated to the validation set. The re-segmentation ensures that the model encounters plasma current magnitudes in the validation set that it has not been exposed to during training. Figure \ref{fig:outinference2} illustrates how the coefficient of determination ($R^2$) for all outputs varies with different intervals of plasma current in the validation set.

The trends shown in figure \ref{fig:outinference2} reveal that the model's predictive accuracy remains relatively consistent for plasma currents up to 900kA. Beyond this threshold, a noticeable decrease in $R^2$ values for certain parameters suggests a decline in the model's extrapolation performance. The observation implies that while the model can extend its predictions to some degree beyond the training range, its ability to do so weakens as the plasma current deviates further from the trained intervals. Besides, the model's predictive capability for $q_{95}$ appears to diminish significantly when the plasma current exceeds 900kA. The decreasing talent on predicting $q_{95}$ would be probably attributed to the intrinsic relationship between $q_{95}$ and plasma current.

\section{Summary}

In this work, a neural network model named EFITNN is designed to reconstruct the real-time magnetic equilibrium for HL-3 tokamak, where eight essential plasma parameters, the poloidal magnetic flux and toroidal current density distributions with a high resolution of $129\times129$ can be obtained. The model's accuracy satisfies the stringent demands of real-time plasma control. The dataset comprises 1 515 618 time slices from valid discharges of the HL-3, ranging from shot \#1248 to shot \#4186. The input features for EFITNN are derived from 68 channels of magnetic measurement data, consistent with those utilized by the offline EFIT code. The results show that EFITNN demonstrates robust performance on unseen data, achieving an average $R^2 > 0.94 $ for 8 plasma parameters, $R^2 > 0.99$ for $\Psi_{rz}$ reconstructions, and $R^2 > 0.95$ for $J_\text{P}$ reconstructions.

In terms of computational efficiency, EFITNN has an average processing time of 0.08ms (model for predicting eight parameters) to 0.45ms (models for predicting $129\times$129 distributions) per time slice (computed by NVIDIA A100 Tensor Core GPU with TensorRT), which meets the real-time control requirements. EFITNN integrates fully connected, convolutional, and deconvolutional network layers, with the latter two particularly adept at identifying and reconstructing local features within the data. Their local-connected feature not only offers an advantage over models relying solely on fully connected networks, but also makes more effective use of neurons, and thereby decreases the complexity of the network.

EFITNN's architecture is further enhanced by a multi-task learning structure. The architecture enables outputs governed by the same physical laws to provide mutual corrective feedback, significantly boosting the model's predictive accuracy. The model also exhibits commendable generalization capabilities, as evidenced by its successful predictions of QSF divertor configurations, among other tests.

EFITNN currently replicates only a subset of the functionalities provided by offline EFIT, there is ample scope for expansion. Future iterations of the model could incorporate additional outputs vital for real-time control, such as the safety factor profile and the location of the X-point. Moreover, integrating boundary conditions or other constraints from offline EFIT could further refine its accuracy. The continued development of EFITNN promises to enhance its utility as a tool for real-time plasma control in fusion plasmas.

\setcounter{figure}{0}
\renewcommand{\thefigure}{B\arabic{figure}}
\begin{figure*}[!t]
    \centering
    \includegraphics[width=\textwidth]{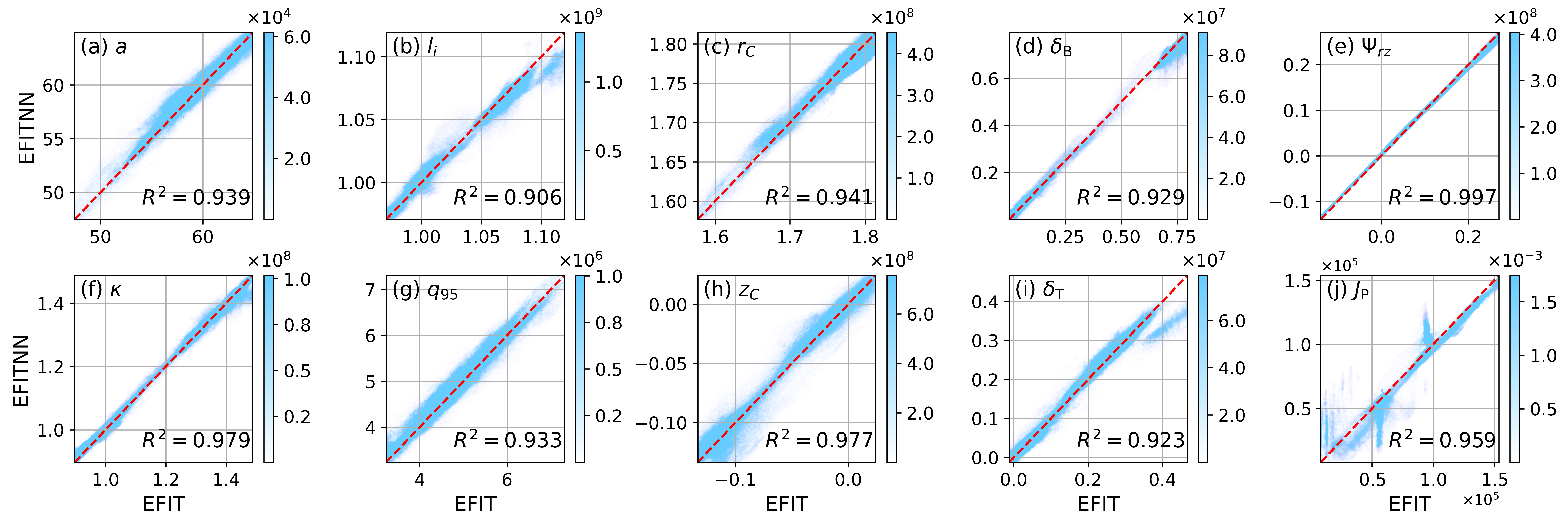}
    \caption{Comparison of values between offline EFIT and EFITNN for 8 plasma parameters, $\Psi_{rz}$ and $J_{\text P}$ from validation and test sets (Shot \#3309-\#4186). The red dashed line represents the $y=x$ line, and the color of the dots indicates the density of data points in the sample space. Each dot represents the value from a time slice, and dots shown in $\Psi_{rz}$ and $J_\text{P}$ are the average of the $129\times 129$ values.}
    \label{fig:vs}
\end{figure*}

\appendix
\section{Detailed parameters in network layers}

\setcounter{table}{0}
\renewcommand{\thetable}{A\arabic{table}}

The appendix provides a comprehensive breakdown of the hyperparameters for the branch network layers within EFITNN. While the hyperparameters for the shared layers and the layers dedicated to predicting plasma parameters (para layers) are detailed in table \ref{tab:config}, the network's architecture must be adjusted when the primary objective shifts to predict the scalar field distributions of $\Psi_{rz}$ and $J_{\text P}$. The adjustments are necessary to accommodate the different resolutions required for the predictions.

Tables \ref{tab:out1} and \ref{tab:out2} enumerate the specific hyperparameters used in the branch network layers when it focuses on primarily predicting $\Psi_{rz}$ and $J_{\text P}$, respectively. The parameters include the number of layers, the number of filters per layer, kernel sizes, strides, and any other relevant architectural details that pertain to the convolutional and deconvolutional layers involved in processing the scalar field outputs. The details are crucial for replicating the network's structure and ensuring the accurate reconstruction of the plasma's magnetic properties at the desired resolutions.

\begin{table}[H]
    \centering
    \caption{Hyperparameters for the branches outputting 17$\times$17 scalar field distributions when the primary objective is the prediction of plasma parameters (the hyperparameters are identical for branches outputting $\Psi_{rz}$ and $J_{\text P}$).}
    \scalebox{0.9}{
        \begin{tabular}{cccccc}
            \Xhline{1pt}
            \Gape[8pt]{Type} & Stride & Kernel size & Filters & Paddings & Regularizations \\
            \hline
            \Gape[6pt]{DeConv2d} & (2, 2) & (3, 3) & 128 & None & BN \\
            \Gape[6pt]{DeConv2d} & (2, 2) & (3, 3) & 128 & None & BN \\
            \Gape[6pt]{DeConv2d} & (2, 2) & (3, 3) & 128 & None & BN \\
            \Gape[6pt]{DeConv2d} & (1, 1) & (3, 3) & 64 & (2, 2) & BN \\
            \Gape[6pt]{Conv2d} & (1, 1) & (3, 3) & 64 & None & BN \\
            \Gape[6pt]{Conv2d} & (1, 1) & (3, 3) & 64 & None & Dropout \\
            \Xhline{1pt}
        \end{tabular}
    }
    \label{tab:out1}
\end{table}

\begin{table}[H]
    \centering
    \caption{Hyperparameters for the branch predicting 129$\times$129 scalar field distributions when the primary objective is the prediction of $\Psi_{rz}$ or $J_{\text P}$ (the hyperparameters are consistent for them).}
    \scalebox{0.9}{
        \begin{tabular}{cccccc}
            \Xhline{1pt}
            \Gape[8pt]{Type} & Stride & Kernel size & Filters & Paddings & Regularizations \\
            \hline
            \Gape[6pt]{DeConv2d} & (4, 4) & (7, 7) & 128 & None & BN \\
            \Gape[6pt]{DeConv2d} & (4, 4) & (7, 7) & 128 & None & BN \\
            \Gape[6pt]{DeConv2d} & (4, 4) & (7, 7) & 128 & None & BN \\
            \Gape[6pt]{DeConv2d} & (2, 2) & (3, 3) & 64 & (3, 3) & BN \\
            \Gape[6pt]{Conv2d} & (1, 1) & (3, 3) & 64 & None & BN \\
            \Gape[6pt]{Conv2d} & (1, 1) & (3, 3) & 64 & None & Dropout \\
            \Xhline{1pt}
        \end{tabular}
    }
    \label{tab:out2}
\end{table}

For tables listed above, Conv2d and DeConv2d represent convolutional and deconvolutional layers, while BN and Dropout indicate the regularization mechanisms attached after each layer, which match batch normalization and dropout, respectively.

\quad

\section{Results from EFITNN vs. offline EFIT}

This appendix displays correlation figures for results from EFITNN and offline EFIT, which has been discussed in Section 4.1. Figure \ref{fig:vs} depicts a comparison of results between them for all EFITNN outputs, it demonstrates good agreement for them.

\quad

\section*{Acknowledgments}
This work is supported by National Natural Science Foundation of China under grant No. 12275142 \& No. U21A20440, and Artificial Intelligence Key Laboratory of Sichuan Province under grant No. 2023RZY03, as well as Natural Science Foundation of Sichuan Province under grant No. 2023NSFSC1291. The authors would like to thank the HL-2A \& HL-3 collaboration groups and the CODIS team for their invaluable support.

\bibliographystyle{unsrt}
\bibliography{ref}

\end{document}